\documentstyle[editedvolume]{crckapb}

\begin{opening}
\title{THE A\&A EXPERIENCE WITH IMPACT FACTORS}
\author{Aa. SANDQVIST}
\institute{Stockholm Observatory\\
                SCFAB-AlbaNova, SE-106 91 Stockholm, Sweden}
\end{opening}

\begin{document}

\section{Introduction}

There is a widespread impression that the scientific journal {\it
Astronomy \& Astrophysics} (A\&A) has a smaller impact, as measured by
citations to articles, than some of the other major astronomy journals.
This impression is apparently supported {\it and probably created} by the
Journal Citation Report (JCR), which is prepared annually by the
Institute of Scientific Information (ISI) Web of Knowledge. In the JCR
statistics, A\&A has over a number of years shown a considerably lower
impact factor than some of the other main journals in the field. For
example, Abt (2003) presented a table in last year's volume of this
series (Organizations and Strategies of Astronomy - Vol. 4) in which A\&A
was in 15th place (behind such journals as "Acta Astron." and "Astrophys.
Lett. \& Comm."). The published poor impact factor of A\&A is in fact due
to a serious flaw in the method used by ISI Web of Knowledge to determine
it. The resulting damage inflicted upon A\&A by the JCR is
incalculable. Attempts to correct the wrong impression are now
proceeding (Sandqvist 2003, Heck 2003).

\section{JCR's Erroneous Method and Its Correction}

Owing to the short abbreviation (A\&A) with which articles in {\it
Astronomy \& Astrophysics} are usually cited, and the possible
non-uniqueness of this abbreviation among the scientific journals covered
by the JCR, these were not counted. Instead, such old abbreviations as
"Astron. Astrophys." were used by JCR. Anyone active in the astronomical
literature will realize the catastrophic effect of this choice. As
noted by Abt (2003), a símilar situation occurred with The
Astrophysical Journal, for which the three-letter abbreviation (ApJ)
had been considered too challenging. After accounting for citations
with ApJ, the impact factor of The Astrophysical Journal increased by
more than a factor of two from 2000 to 2001.

As chairman of the A\&A Board of Directors, I contacted the ISI Web of
Knowledge in the fall of 2002, as soon as we had discovered the erroneous
abbreviation that JCR was using for A\&A. After some hesitation, the ISI
Web of Knowledge decided to change its routines, and in the future the
abbreviation A\&A would be taken into account. In-house studies at the
ISI Web of Knowledge showed that very little confusion arose through
this, and more importantly, after accounting for these citations, the
impact factor of A\&A became comparable to that of the other major
astronomy journals and the total number of citations second only to The
Astrophysical Journal. Apologizing, the ISI informed me in the fall of
2003 that it had made a reconstruction of  what A\&A's impact factor
would have been in the year 2000, if the proper Journal abbreviation had
been used. The result was heartening for A\&A: the A\&A impact factor had
rebounded from a wrong value of 2.79 to a corrected value of 4.352 with
a resulting change of rank from a wrong value of 11th to a corrected
value of 4th out of 37 in the category of Astronomy and Astrophysics!

Since the impact factor is defined as the number of papers cited for a
given journal divided by the total number of papers published in that
journal, averaged over the past two years, the true positive effect of
this correction for A\&A will be slower than desired.

\section{Confidence in ISI?}

How trustworthy are the citation statistics published by JCR? Given that
ISI is a commercial enterprise, expensive, and that evaluations of
individuals, institutes and scientific disciplines are based on their
data, their consumers have a right to expect absolute professionalism. We
have seen one example of gross failure in the previous section. Another
example (which, however, does not involve A\&A directly) is detailed
below.

On ISI, a January 2004 search for citations of  SCHNEIDER P in 2001
yielded:

{\it 2      Schneider P.      PHYS REP     340      292      2001}\\
where "2" is the total number of citations.\\
A similar search on the NASA Astrophysics Data System (ADS) yielded:

{\it 2001 PhR   340 ...291B   113.000    01/2001    Bartelmann, M.;
Schneider, P.     Weak gravitational lensing}\\  
where "113" is the total number of citations.

This huge difference in the total number of citations reported by the ISI
and ADS is due to an error in the ISI data base which claimed the first
page number of the article to be "292", whereas its correct value should
be "291" as properly stated by ADS. A search using the name of the first
author (Bartelmann) did give the correct number of citations for this
article at both ISI and ADS, but this does not help Dr. Schneider,
though. When informed of this error, the ISI responded that a correction
could take "up to 4 weeks". This reply does not show the level of
professionalism that one would expect from a company which has taken a
large influence in shaping the careers of scientists. Whether the
astronomical funding agencies should continue to rely heavily upon ISI
when other citation tools have become available now appears questionable.

\section{Damage Inflicted Upon A\&A}

Although the A\&A Board of Directors is gratified by the new development
in the JCR with respect to A\&A, we are extremely concerned about the
damage inflicted upon our Journal by ISI, in particular since it also has
had detrimental personal and institutional consequences for A\&A's
authors. Also serious is the fact that it will take some time before the
full impact of this change in the ISI software will have been made, so
that it correctly reflects the proper citation index and impact factor
for our Journal. And how long will it take before the Astronomical
Community has been made fully aware of the JCR error and its correction,
not to mention the difficulty of reaching financing agencies which make
ample use of journal impact factors?

Problems that have come to our attention vary in different countries:

(1) Some libraries are considering cancelling subscriptions to A\&A since
they only want the astronomical journals with the ten highest impact
factors. The erroneous published value of 15th place for A\&A would
thus disqualify it for these libraries' acquisition lists. 

(2) Some ministries naturally turn to impact factors when making
comparative studies between scientific disciplines and their general
impact, with the purpose of drawing up priority lists for future
large-scale financing. Such studies, making use of the erroneous
published values, will have had very serious effects upon the efforts of
some A\&A authors to achieve future financial support.

(3) Some institutes do not select A\&A papers in their top ten list in
their annual reports due to the erroneous published value of the A\&A
impact factor.

(4) Some astronomers use the low erroneous published value for the A\&A
impact factor as a reason for not publishing in A\&A which, of course, is
highly detrimental to our Journal.

\section{Conclusion}

The damage to A\&A has been done. The error has been found. The error has
been corrected by ISI. This information must now be disseminated
throughout the Astronomical Community, the Institutional Libraries, the
Financing Agencies, the General Society of Impact Factor Consumers. This
short communication is one such attempt. The A\&A Board of Directors is
grateful if you, the reader, will do the same.

\section{References}

Abt, H.A. (2003) The Institute for Scientific Information and the
Science Citation Index, in {\it Organizations and Strategies in
Astronomy - Vol. 4}, Ed. A. Heck, Kluwer Academic  Publishers,
Dordrecht, pp.~197-204\\ 

\noindent Heck, A. (2003) WRONG IMPACT!, {\it European Astronomical Society
Newsletter}, {\bf Issue 26}, December 2003, pp.~4-5\\

\noindent Sandqvist, Aa. (2003) Remark on Impact Factor, {\it A\&A}, {\bf
Vol.~no.~402}, p.~E1\\\\\\\\\\\\\\\\\\\\\\\\\\\\\\\\\\\\\\\\

{\it To be published in ``Organizations and Strategies in Astronomy --
Vol. 5'', Ed. A. Heck, Kluwer Academic  Publishers, Dordrecht (2004)}

\end{document}